\documentclass[twocolumn,showpacs,preprintnumbers,amsmath,amssymb,prl,aps]{revtex4-1}
\usepackage{graphicx}
\usepackage{bm}
\usepackage{float} 
\usepackage{amsmath}
\usepackage{gensymb}
\usepackage{epstopdf}
\usepackage[usenames, dvipsnames]{color}

\begin{document}

\title{Anomalous ground state properties of SmB$_6$ - a density functional theoretical study}

\author{Anup Pradhan Sakhya}
\author{Kalobaran Maiti}
\altaffiliation{Corresponding author: kbmaiti@tifr.res.in}

\affiliation{Department of Condensed Matter Physics and Materials Science, Tata Institute of Fundamental Research, Homi Bhabha Road, Colaba, Mumbai - 400 005, INDIA.}

\begin{abstract}
We studied the electronic structure of SmB$_6$ employing density functional theory (DFT) using different exchange potentials, spin-orbit coupling (SOC) and electron correlation ($U$ = correlation strength). All the calculations carried out in this study converge to metallic ground state indicating bulk metallicity in SmB$_6$, which is in line with the low temperature anomalies observed in this system. We show that while spin-orbit coupling and electron correlation is important to capture the ground state properties, generalized gradient approximation provides the best description of the Sm 4$f$ multiplets observed in angle-resolved photoemission spectroscopy (ARPES) data. The Fermi surface plots exhibit electron pockets around $X$-point and hole pockets around $\Gamma X$ line having dominant Sm 4$f$ character; the observation of these Fermi surfaces is consistent with the recent quantum oscillation measurements. In addition to primarily Sm 4$f$ contributions observed at the Fermi level, we discover significantly large contribution from B 2$p$ states compared to weak Sm 5$d$ contributions. This suggests important role of B 2$p$ - Sm 4$f$ hybridization in the exotic physics of this system.
\end{abstract}

\date{\today}


\maketitle

\section{Introduction}

Mixed valent Kondo insulators have attracted tremendous attention followed by the discovery of varied exotic ground state properties arising from strong Coulomb repulsion among 4$f$ electrons and their hybridization to conduction electrons \cite{Fisk,CeRhCoSi3,Rise,Cole,Anto}. SmB$_6$ is one such material \cite{Fisk,Rise,Anderson} exhibits metallicity at room temperature and becomes insulating below 40 K, which is believed to arise due to the formation of many body singlet state constituted by localized Sm 4$f$ electrons and the dispersive Sm 5$d$ electrons. Various experimental studies revealed anomalies at low temperatures such as finite linear specific heat coefficient, bulk optical conductivity below the charge gap, quantum oscillations within the insulating phase, and saturation of resistance below 4 K, which was attributed to the presence of {\it in-gap} states within the charge gap \cite{Menth}.

First principles calculations of the electronic structure of SmB$_6$ predicted non-trivial $Z_2$ topology that may host topologically protected metallic surface states leading to a saturation of resistance at low temperatures \cite{Coleman,Galitski}. Subsequent transport \cite{Wolgast,Kim,DJKim} and angle-resolved photoemission spectroscopy (ARPES) \cite{Neupane,Jiang,NXu,Fran} measurements on SmB$_6$ supports the presence of topologically ordered metallic surface states. Employing spin-resolved ARPES measurements, Xu {\it et al.} \cite{sparpes} showed that the metallic surface states in this material are spin polarized and the spin texture fulfills the condition that the surface states are protected by time-reversal symmetry making SmB$_6$ an example of a topological Kondo insulator.

Quantum oscillation experiments by Li {\it et al.} exhibit signature of two dimensional Fermi surfaces on (100) and (101) surface planes supporting the presence of topological surface states within the bulk hybridization gap \cite{GLi}. Recent studies, however, argued that the observed metallic surface states has trivial origin rendering SmB$_6$ a trivial surface conductor \cite{Radar}. Torque magnetometry experiments by Tan {\it et al.} \cite{Tan} exhibit angular dependence of de Haas van Alphen oscillations suggesting three dimensional nature of the observed Fermi surfaces. It was suggested that the high frequency quantum oscillations originate from a large three dimensional Fermi surface occupying half the Brillouin zone, which strongly resemble the $d$ type conduction electron Fermi surface in metallic LaB$_6$ \cite{Tan,Hartstein}. In order to explain low temperature anomalies, some groups proposed Fermi surfaces due to neutral fermionic composite exciton \cite{baskaran,senthil}. Recently, Harrison {\it et al.} \cite{harrison} has shown the presence of highly asymmetric nodal semi-metal phase existing over certain region of momentum space in bulk SmB$_6$, where the node is pinned to the un-hybridized $f$-level, casting doubt over the necessity of a neutral Fermi surface.

Evidently, SmB$_6$ is a novel material exhibiting several outstanding puzzles in its electronic properties. We have calculated the electronic structure of SmB$_6$ employing density functional theory (DFT) using various exchange correlation potentials. We discover that the ground state of the bulk electronic structure of SmB$_6$ is metallic and the metallicity survives even if we change the exchange correlation potentials. The results obtained considering electron correlation and spin-orbit coupling within the DFT level capture the ARPES results as well as other bulk electronic properties of this system.

\section{Computational Details}

The electronic structure calculations were performed using the full-potential linearized augmented plane-wave (FLAPW) method as implemented in the WIEN2k software \cite{Wien}. We have used the generalized-gradient approximation (GGA) for the exchange correlation functional proposed by Perdew-Burke-Ernzerhof (PBE) \cite{Perdew1,Perdew2}, where the functional depends on local charge density as well as on the spatial variation of the charge density. In order to verify the sensitivity of the results on the choice of exchange correlation potentials, we have also calculated the electronic structure using modified Becke Johnson (mBJ) potential \cite{tran}, where the potential uses information from kinetic energy density in addition to the charge density. The mBJ potential corresponds to an orbital-independent semi-local exchange potential mimicking the orbital dependent behavior. It has been found to yield good description of the band gaps, effective masses and correct band ordering at time-reversal invariant momenta (TRIM), and are in good agreement with the improved many-body but more computation demanding GW (G = Green's function and W = screened Coulomb interaction) calculations \cite{singh,Marsman,Zhong}. Calculations were performed with and without inclusion of electron correlation ($U$ = electron-electron Coulomb repulsion strength) and spin-orbit coupling (SOC). From the experiments, it is observed that SmB$_6$ does not show magnetic order down to 19 mK \cite{Biswas2}. In order to simulate this, we have performed constrained magnetic calculations so that the magnetism of SmB$_6$ is consistent with the experimental scenario.

\begin{figure}
\includegraphics[scale=.25]{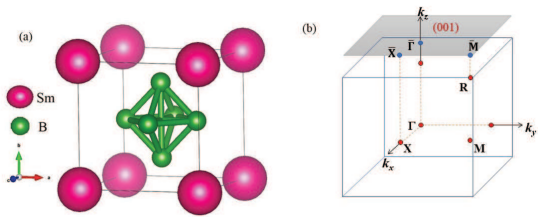}
\caption{(color online) (a) Crystal structure of SmB$_6$ exhibiting simple cubic symmetry. (b) The bulk and surface Brillouin zones.}
\label{struct}
\end{figure}

SmB$_6$ forms in cubic structure with Sm atoms at the corner of the cube and B$_6$ octahedra at the body center as shown in Fig. \ref{struct}(a). The crystal structure possesses inversion symmetry. We have used experimentally observed lattice constants of SmB$_6$ \cite{smb6latt} for our calculations. In Fig. \ref{struct}(b), we show the bulk Brillouin zone (BZ) having cubic symmetry; the centre of the BZ is the $\Gamma$ point and the edge centre, face center \& corner are denoted by $M$, $X$ \& $R$ points, respectively. The shaded area on top represents the projection of the bulk BZ at the surface. The high symmetry points on the surface BZ are represented by $\overline{\Gamma}$, $\overline{X}$ and $\overline{M}$. The energy bands were calculated along various $k$-vectors shown in the figure. The Fermi surfaces were calculated using Xcrysden with 31$\times$31$\times$31 $k$ mesh. The spin-orbit coupling was included self-consistently in the electronic structure calculations with a 17$\times$17$\times$17 $k$-mesh.


\section{Results and Discussions}

\begin{figure}
\includegraphics[scale=0.5]{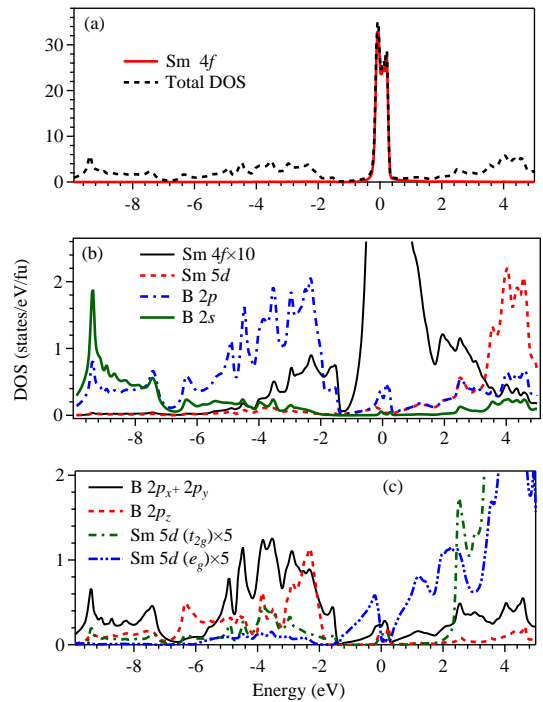}
\vspace{-2ex}
\caption{(color online) Calculated (a) total density of states (TDOS) (dashed line) and Sm 4$f$ partial density states (PDOS, solid line). (b) PDOS of Sm 4$f$ (rescaled by 10 times - solid line), Sm 5$d$ (dashed line), B 2$p$ (dot-dashed line) and B 2$s$ (green solid line). (c) B 2$p_x$+2$p_y$ PDOS (solid line), B 2$p_z$ PDOS (dashed line), and 5 times of Sm 5$d$ PDOS with $t_{2g}$ (dot-dashed line) and $e_g$ (dot-dot-dashed line).}
\label{ggados}
\end{figure}


In Fig. \ref{ggados}(a), we show the calculated total density of states (TDOS) obtained using GGA for the exchange correlation potential. Partial density of states (PDOS) have been calculated by projecting the eigenstates onto the atomic orbitals. The calculations converged to a metallic ground state with huge intensity at the Fermi level, $E_F$ as evident in the figure. The region near $E_F$ (- 0.3 eV to 0.4 eV) is contributed mostly by Sm 4$f$ states and the lower energy part of the valence band (VB) region (-6 eV to -1.3 eV) is dominated by contributions from B 2$p$ states with some contribution from B 2$s$ states as shown in Fig. \ref{ggados}(b) and \ref{ggados}(c). The B 2$s$ PDOS predominantly appear in the energy range from  -7 eV to -10 eV and play little role in deriving the electronic properties of this material.

In Fig. \ref{ggados}(b), we show Sm 5$d$, B 2$p$ contributions along with rescaled Sm 4$f$ PDOS to compare the energy distribution of the density of states. It is evident that the energy distribution of the Sm 4$f$ contributions look similar to B 2$p$ contributions, which is a signature of hybridization between Sm 4$f$ - B 2$p$ states. The bonding bands appear between -7 eV to -1.6 eV with a peak at about -2.5 eV and the anti-bonding bands appear above -1.6 eV. Sm 5$d$ states are essentially unoccupied and contribute beyond 3 eV above $E_F$. There are weak Sm 5$d$ contributions within the valence band regime suggesting finite coupling of Sm 5$d$ - B 2$p$ states. Interestingly, the Sm 5$d$ PDOS observed here look very similar to the $d$ PDOS in other hexaborides such as LaB$_6$, CaB$_6$ \cite{LaB6APL,CaB6EPL} where 4$f$ contributions are not present. This suggests that $d$-$p$ hybridizations are quite similar in this class of materials and 4$f$ states may not be influencing the $d$ states significantly. The energy distribution of various PDOS also manifests similar scenario.

In the crystal structure of SmB$_6$, Sm ions at each corner is surrounded by eight B$_6$ clusters located at the center of the cube and hence, Sm sites will experience cubic crystal field. The Sm 5$d$ orbitals have larger radial extensions and hence, will experience strong crystal field effect. The cubic crystal field splits the Sm 5$d$ levels into a doubly degenerate $e_g$ band and a triply degenerate $t_{2g}$ band. For the axis system similar to the crystal lattice axis and boron clusters located at the center of the cube, $e_g$ states will have orbital lobes away from the clusters and hence, the energy for $e_g$ electrons will be lower (weaker Coulomb repulsion energy) than the $t_{2g}$ electrons possessing orbital lobes along the anion clusters due to repulsion of $d$ electrons due to negative ligand charges \cite{Kang}. Thus, Sm 5$d$ ($e_g$) states are partially occupied and contribute at $E_F$; Sm 5$d$ ($t_{2g}$) bands are essentially unoccupied lying high in energy (above 2.4 eV) in the conduction band as shown in Fig. \ref{ggados}(c). B (2$p_x$+2$p_y$) PDOS shown in Fig. \ref{ggados}(c) exhibit similar energy distribution of Sm 4$f$ PDOS indicating stronger hybridization with these states. B 2$p_z$ seem to have significant hybridization with the Sm 5$d$ states. From the PDOS plots, we find that the major contribution to the total DOS at $E_F$ [$N(E_F)$] comes from the Sm 4$f$ states (98.6 \%), while the contribution from the Sm 5$d$ and B 2$p$ states are 0.16 \% and 1.1\%, respectively. The contribution from B (2$p_x$ + 2$p_y$) is more near $E_F$ than B 2$p_z$ states.

\begin{figure}[h]
   \centering
  \includegraphics[scale=.5]{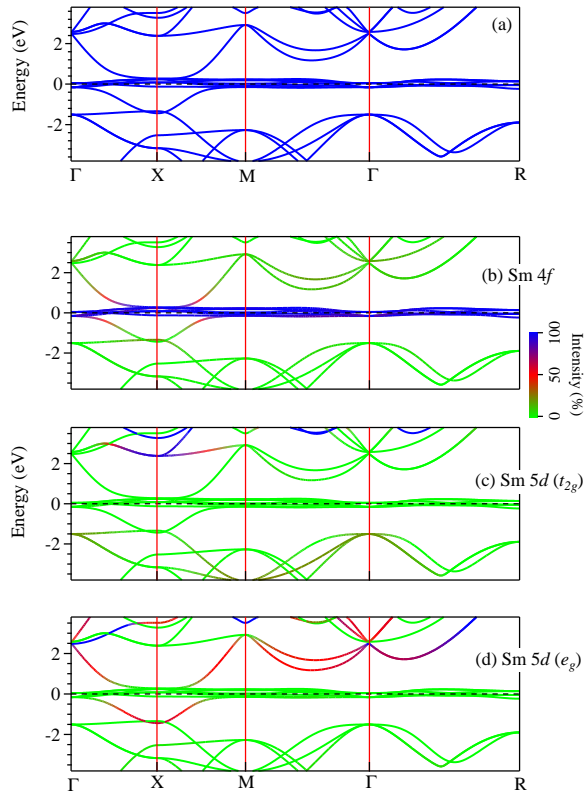}
   \vspace{-1ex}
     \caption{(color online) (a) Energy band structure obtained using GGA along various $k$-vectors. Contribution of (b) Sm 4$f$, (c) Sm 5$d$ ($t_{2g}$), and (d) Sm 5$d$ ($e_g$) to various energy bands are shown [Sm 5$d$ contributions are rescaled by a factor of 2 for clarity].}
     \label{smbgga}
\end{figure}

The calculated energy band structure along high symmetry directions are shown in Fig. \ref{smbgga}. While the DOS provide information about the contribution of various electronic states as a function of energy ($k$-integrated results), the band structure provides the $k$-resolved information that can be compared directly with the data from angle resolved photoemission spectroscopy (ARPES) measurements. In order to get more information regarding the contribution from different states, we have plotted the band structure with band-character plots. The energy bands near $E_F$ exhibit minimal dispersion indicating high degree of local character of the corresponding electronic states. From the colour plots in Fig. \ref{smbgga}(b), it is clear that these valence states possess essentially Sm 4$f$ character as also manifested in the PDOS plots shown in Fig. \ref{ggados}.

A highly dispersive energy band crosses the Fermi level along $\Gamma X$ and $XM$ directions, and hybridizes with the Sm 4$f$ bands. From the symmetry analysis shown in Figs. \ref{smbgga}(c) and \ref{smbgga}(d), we observe that the bands having $e_g$ symmetry hybridizes strongly with the Sm 4$f$ bands; $t_{2g}$ bands appear at higher energies. The projected band characters provide signature of band inversion near the Fermi level. In addition, at about -1.5 eV, two bands touch each other at $X$-point.
Transport \cite{Menth,Nickerson,Cooley} and optical measurements \cite{Molnar,Travaglini,Demsar} exhibit signature of a small gap at temperatures below 50 K, which is in contrast to the calculated metallic ground state found here. This suggests the necessity to go beyond GGA. Since, Sm is a heavy element ($Z$ = 62), spin orbit coupling (SOC) is expected to play an important role in the electronic properties of this material ($\Delta \varpropto Z^4$). As a first step, we calculated the electronic structure including SOC and different exchange correlation potentials.

\begin{figure}
\centering
  \includegraphics[scale=.5]{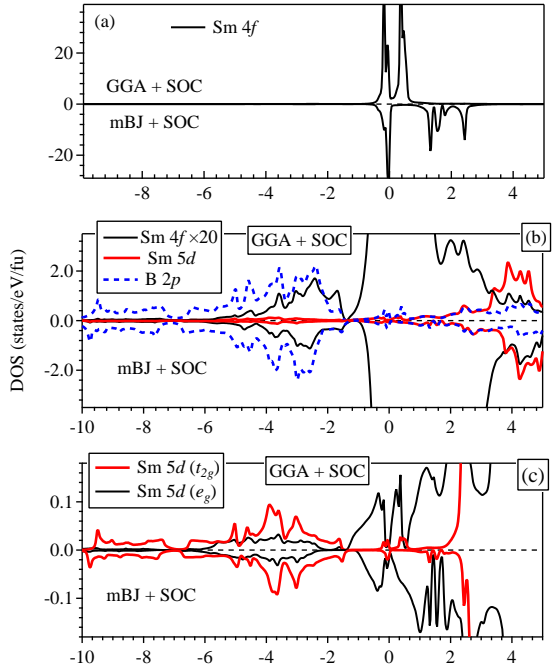}
    \vspace{-2ex}
  \caption{(color online) Calculated partial density of states of (a) Sm 4$f$, (b) Sm 4$f$ (rescaled by 20 times shown by thin black solid lines), Sm 5$d$ (thick red solid lines), B 2$p$ (blue dashed lines), and (c) Sm 5$d$ ($t_{2g}$) (thick red solid lines) and Sm 5$d$ ($e_g$) (thin black solid lines) states. GGA+SOC results are shown along positive $y$-axis and mBJ+SOC results are shown along negative axis for better comparison.}
  \label{socdos}
\end{figure}

The DOS and PDOS calculated with the inclusion of spin-orbit coupling is shown in Fig. \ref{socdos}(a) exhibiting significant changes near $E_F$. The Sm 4$f$ bands split into Sm 4$f_{5/2}$ ($J$ = 5/2) and Sm 4$f_{7/2}$ ($J$ = 7/2) levels. The Sm 4$f_{5/2}$ band is partially occupied and lie in the energy range -0.4 eV to 0.1 eV, while Sm 4$f_{7/2}$ band is empty and lie in the energy range 0.2 eV to 0.7 eV. In addition, there is a marginal decrease in Sm 4$f$ contribution at $E_F$ (from 98.6\% to 97.5\%) with consequent increase in the Sm 5$d$ and B 2$p$ contributions to 0.3\% and 2\%, respectively. The ground state remains metallic in these results with significant DOS at $E_F$ although a pseudogap like feature (peak-dip-peak structure) appears at $E_F$.

In order to verify the effect of the approximations in exchange correlation potentials in the electronic structure, we calculated the electronic structure using modified Becke Johnson potential including SOC (mBJ+SOC) method; the results are shown in Fig. \ref{socdos}(a) with reverse axis direction for better comparison. It appears that the DOS at $E_F$ is slightly less than the results from GGA+SOC calculations but there is still substantial DOS at $E_F$. The pseudogap like feature observed in GGA+SOC results survives along with a high degree of particle-hole asymmetry. The Sm 4$f$, 5$d$ and B 2$p$ PDOS below $E_F$ exhibit signature of strong covalency between Sm 5$d$-B 2$p$ and Sm 4$f$-B 2$p$ states. $N(E_F)$ in mBJ+SOC results is composed primarily of Sm 4$f$ states contributing 98.7\%, with the second most prominent contribution from B 2$p$ states (1\%), while the contribution from Sm 5$d$ states is only 0.1\%.

\begin{figure}
   \centering
   \includegraphics[scale=.5]{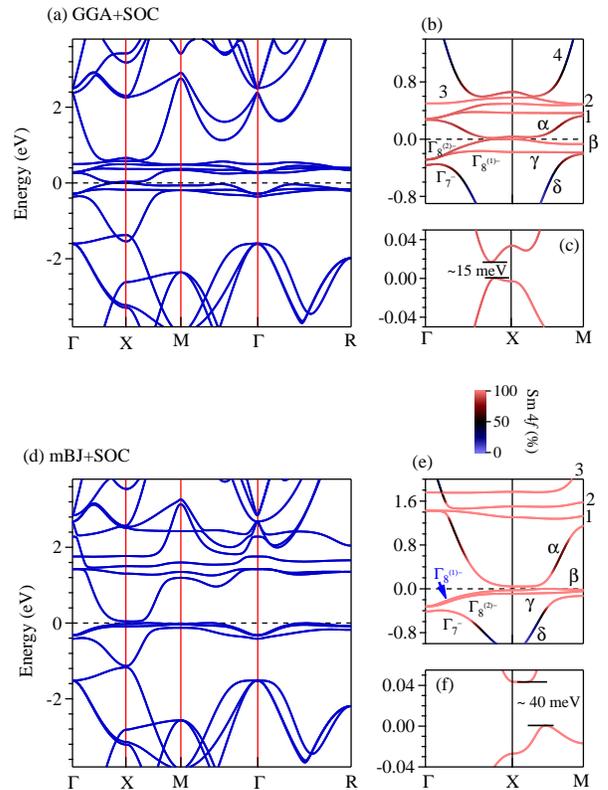}
   \vspace{-1ex}
     \caption{(color online) Energy band structure of SmB$_6$ using (a) GGA+SOC and (d) mBJ+SOC. The energy region close to $E_F$ along $\Gamma XM$ direction is shown in (b) for GGA+SOC and (e) for mBJ+SOC exhibiting signature of band inversion. Band gaps are shown in (c) for GGA+SOC and (f) mBJ+SOC. Color scale provide relative contribution of various electronic states to the energy bands.}
     \label{bandsoc}
\end{figure}

The band structure along with the band character plots are shown in Fig. \ref{bandsoc}. The inclusion of SOC leads to a splitting of the Sm 4$f$ bands by about 0.6 eV in the GGA+SOC calculations shown in Fig. \ref{bandsoc}(a). In addition, a direct band gap of about 15 meV opens up along $\Gamma-X$ direction (see Fig. \ref{bandsoc}(c)) consistent with earlier results \cite{Dai}. Based on ARPES measurements, Frantzeskakis {\it et al.} \cite{Fran} proposed that the states at the $X$ point are essentially bulk Sm 5$d$ states and the Fermi level lies at about 20 meV below the top of the valence band. The estimated band gap from their experimental results found to be within about 5-10 meV along $XM$ and $\Gamma$-$X$ directions, which is consistent with our theoretical results. We find that the bands exhibiting the gap possess primarily Sm 4$f$ character. Sm 4$f$ bands cross $E_F$ along the $\Gamma-X$ direction and form tiny hole pockets and along $XM$ line, there is an energy gap.

On the other hand, mBJ+SOC results exhibit Fermi level crossing along the $XM$ line; the band gap in this case becomes indirect with an energy gap of about 40 meV, which is much larger than the ARPES results. There exists a tiny hole pocket around $XM$ line with no Fermi surface along $\Gamma X$ direction. The spin-orbit coupling leads to splitting of Sm 4$f$ bands to 4$f_{5/2}$ \& 4$f_{7/2}$ bands with a large SOC splitting of about 1.6 eV. Moreover, the multiplet splitting of 4$f_{7/2}$ bands (band numbers 1, 2, 3 and 4 in the figure) in mBJ+SOC bands is significantly higher than GGA+SOC results.

In both the cases (GGA \& mBJ), the band structure near the Fermi energy is dominated by Sm 4$f_{5/2}$ states as also found in LDA+Gutzwiller method \cite{Dai}. Both GGA+SOC and mBJ+SOC results exhibit signature of band inversion along the high symmetry directions, $\Gamma X$ and $XM$ (see Figs. \ref{bandsoc}(b) and \ref{bandsoc}(e)). This band inversion has been discussed by many authors and is a prerequisite for SmB$_6$ to be a topological insulator. In Fig. \ref{bandsoc}(b) and Fig. \ref{bandsoc}(e), we have denoted the bands as $\alpha$, $\beta$, $\gamma$, $\delta$, 1, 2, 3 and 4. In the vicinity of $X$ point, $\alpha$, $\beta$ and $\gamma$ bands have $f$ orbital character, and the band $\delta$ has $d$ orbital character. There are additional four conduction bands denoted as 1, 2, 3 and 4 representing the signature of spin-orbit split 4$f_{7/2}$. While in GGA+SOC results, these four bands lie very close to $E_F$, they are shifted above 1.4 eV in the mBJ+SOC results (the band 4 is shifted to 2.3 eV above $E_F$ and not shown in the figure). Most of the 4$f$ bands in mBJ+SOC results are not displaying significant dispersion implying strong atomic nature of Sm 4$f$ electrons similar to the LDA+Gutzwiller results \cite{allen,schmidt,Dai}.

In the GGA+SOC and mBJ+SOC data shown above, although the band inversion takes place between $\alpha$ band having negative parity and $\delta$ band having positive parity, the band gap appears between $\alpha$ and $\beta$ bands, i.e., the two 4$f$ bands having the same parity. Thus, the band inversion in SmB$_6$ is complex and different from the band inversion observed in the typical topological insulators such as Bi$_2$Se$_3$ and Bi$_2$Te$_3$, where electron correlation is not important \cite{Deep-band}. This makes SmB$_6$ special having exoticity due to interplay between electron correlation and topological order \cite{Junwon}.

Sm 4$f$ states experience stronger spin-orbit coupling than the crystal field effect due to their small orbital extension and screening by 5$d$ electrons. Thus, 4$f$ levels split into 4$f_{5/2}$ and 4$f_{7/2}$ bands with large energy separation due to SOC. The crystal field splits Sm 4$f_{5/2}$ bands into a {$\Gamma_7$} doublet and a {$\Gamma_8$} quartet. Away from the {$\Gamma$} point, the {$\Gamma_8$} quartet is further splits into {$\Gamma_8^1$} and {$\Gamma_8^2$} doublets, which is shown in the Figs. \ref{bandsoc}(b) and \ref{bandsoc}(e), respectively \cite{Kang}. SmB$_6$ possesses inversion symmetry and $Z_2$ topological invariants have been computed via a parity analysis and found to be $Z_2$ = 1 in this case \cite{Bansil,Fu,Kane}. Thus, SmB$_6$ has been predicted to be topologically non-trivial system with an odd number of gapless surface states.

\begin{figure}
  \includegraphics[scale=.5]{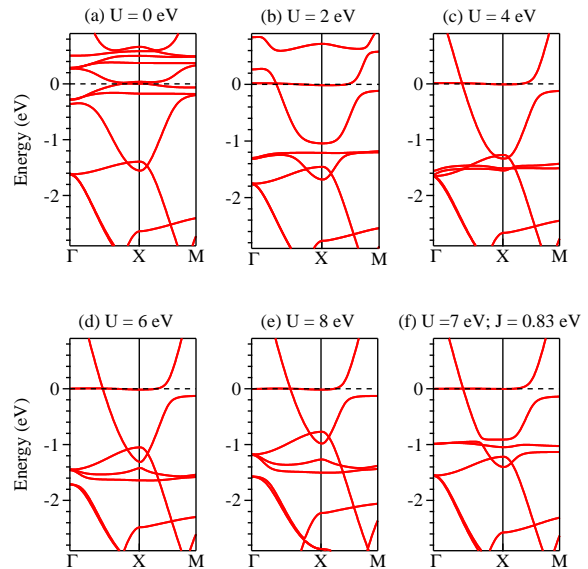}
     \vspace{-1ex}
  \caption{(color online) Energy band structure along $\Gamma-X-M$ direction calculated using GGA+SOC+$U$ method for (a) $U$ = 0, (b) $U$ = 2 eV, (c) $U$ = 4 eV, (d) $U$ = 6 eV, (e) $U$ = 8 eV and $U$ = 7 eV, $J$ = 0.83 eV.}
   \label{GGAplusU}
\end{figure}

While all the above conclusions are interesting and exhibit some features of the experimental observations, the experimental ARPES data \cite{Jiang,Denlinger,Biswas} are significantly different from the results discussed so far. The Sm 4$f$ feature observed at around -1 eV energy (1 eV binding energy in ARPES data) in the experimental data is not found in the calculations performed using GGA, GGA+SOC, mBJ (not shown here) and mBJ+SOC methods. Moreover, the dispersion of the energy bands does not match with the ARPES results. These discrepancies may be related to the underestimation of the electron correlation effects in these methods.

\begin{figure}
 \includegraphics[scale=.5]{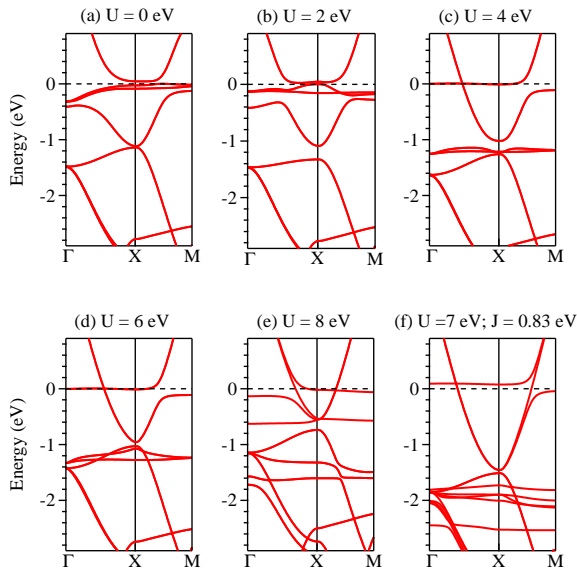}
     \vspace{-1ex}
  \caption{(color online) Energy band structure along $\Gamma-X-M$ direction calculated using mBJ+SOC+$U$ method for (a) $U$ = 0, (b) $U$ = 2 eV, (c) $U$ = 4 eV, (d) $U$ = 6 eV, (e) $U$ = 8 eV and $U$ = 7 eV, $J$ = 0.83 eV.}
   \label{mBJplusU}
\end{figure}

In order to investigate the role of electron correlation on the electronic structure within the density functional theory, we calculated the energy band structure following GGA+SOC+U and mBJ+SOC+U methods ($U$ = electron-electron Coulomb repulsion strength). The results are shown in Figs. \ref{GGAplusU} and \ref{mBJplusU}. We have focused on the $k$-vectors, $\Gamma-X$ and $X-M$ only, where the band cross the Fermi level and the experimental results are available in the literature \cite{Jiang,Denlinger,Biswas}. We have performed the calculations for various values of $U$ ranging from 2 eV to 10 eV for both GGA+SOC+$U$ and mBJ+SOC+$U$.

In Fig. \ref{GGAplusU}, we observe that with the increase in $U$, the 4$f_{5/2}$ bands gradually shift towards higher binding energies. There are significant other changes in the electronic structure due to incorporation of $U$. For example, most of the flat bands near Fermi level in the uncorrelated case shifts to higher energies with the increase in $U$. The energy shifted bands represent the correlation induced bands/lower Hubbard band/incoherent feature. We observe signature of a flat 4$f$ band in the proximity of the Fermi level for all values of $U$ used in our calculations (coherent feature). From the calculations with various combinations of $U$ and $J$, we find that the results for $U$ = 7 eV and $J$ = 0.83 eV provide the best description of the experimental results exhibiting signature of band inversion along $XM$ direction and flat 4$f$ bands at -1 eV. In addition, there is a flat band around -2.5 eV along $XM$ direction.

The results from mBJ+SOC+$U$ calculations appear significantly different from the results of GGA+SOC+$U$ results and experimental data. For example, the data for $U$ = 0 and 2 eV exhibit flat bands representing Sm 4$f$ states near the Fermi level; the change in the band dispersion is found at different energy regimes. For $U$ = 4 eV and 6 eV, the 4$f_{5/2}$ bands appear slightly below -1 eV energy. However, $U$ = 8 eV exhibit flat bands spread over a large energy range leading to complex Sm 4$f$ - B $p$ hybridizations. The $k$-vector exhibiting band inversion scenario depends on the value of $U$ considered in the calculations. While the energetics of the 4$f$ bands for $U$ = 4 eV seems closer to the experiments, the dispersive bands are significantly different from the experimental results. Overall, it was difficult to capture the band structure consistent with experiments using mBJ+SOC+$U$ method.

It is to note here that the Becke-Roussel potential, $v^{BR}_{x,\sigma}{(\bf r)}$ was proposed to model the Coulomb potential due to an exchange hole \cite{BR}, which is very similar to the Slater potential. The mBJ potential was conceived by adding a semi-local correction term to the Becke-Roussel potential to capture the features like step structure and derivative discontinuity of the exchange correlation potential at integral particle number \cite{tran} and is expressed as,

 $$ v^{mBJ}_{x,\sigma}{(\bf r)}=cv^{BR}_{x,\sigma}{(\bf r)}+(3c-2) \frac{1}{\pi}\sqrt{\frac{5}{12}}\sqrt{\frac{2t_{\sigma}(\bf r)}{\rho_{\sigma}(\bf r)}}
$$

where the electron density, ${\rho_{\sigma}}=\sum_{i=1}^{N_\sigma}|{\psi_{i,\sigma}}|^2$, the kinetic-energy density, $t_{\sigma}=(1/2)\sum_{i=1}^{N_{\sigma}}|\nabla\psi_{i,\sigma}|^2$, and $c=\alpha+\beta(\frac{1}{V_{cell}}\int_{cell}\frac{|\nabla{\rho({\bf r')}|}}{\rho (\bf r')}d^{3}r')^{\frac{1}{2}}$. The values of ${\alpha}$ (= -0.012) and ${\beta}$ (= 1.023 (Bohr)$^{1/2}$) are fixed by comparing the theoretical results with the experimental data of a large number of materials \cite{tran}. While this potential is quite successful in predicting the band gap of wide varieties of systems ranging from wide bang gap insulators to correlated transition metal oxides \cite{tran}, polarizabilities in insulators with significant accuracy \cite{Armiento}, our results indicate that GGA calculations provide a better description of the energy band structure in a rare-earth based strongly correlated system such as SmB$_6$. In order to achieve better description, one might require to tune the free parameters further; we hope that these results will provide incentive to initiate such studies in the future.

\begin{figure}
 \includegraphics[scale=.5]{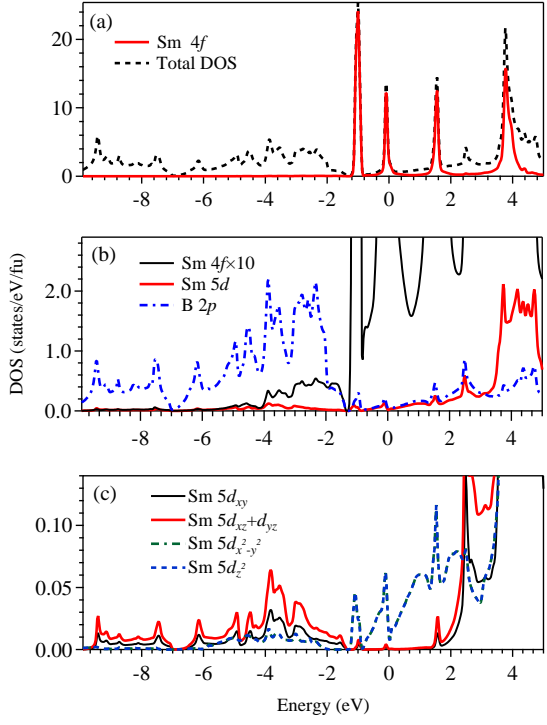}
  \caption{(color online) Calculated results using GGA+SOC+$U$ method ($U$ = 7 eV, $J$ = 0.83 eV). (a) Total density of states (dashed line) and Sm 4$f$ partial density of states (solid line). (b) Rescaled (10 times) Sm 4$f$ PDOS (thin black solid line), Sm 5$d$ PDOS (thick red solid line) and B 2$p$ PDOS (blue dashed line). (c) PDOS of Sm 5$d_{xy}$ (thin black solid line), (5$d_{xz}$ + 5$d_{yz}$) (thick red solid line), Sm 5$d_{x^2-y^2}$ (green dot-dashed line) and Sm 5$d_{z^2}$ (blue dashed line).}
  \label{ggaUeq7dos}
\end{figure}

We now turn to the discussion of the detailed electronic structure for the case, which captures the experimental results well. The DOS obtained using GGA+SOC+$U$ ($U$ = 7 eV and $J$ = 0.83 eV) is shown in Fig. \ref{ggaUeq7dos}. Sm 4$f_{7/2}$ features are observed to be shifted to higher energies, thus, rendering the region near the $E_F$ dominated by Sm 4$f_{5/2}$ features. The DOS at the Fermi level is found to be finite as found in other hexaborides \cite{hexaborides} even after the application of GGA+SOC+$U$; there are significant Sm 4$f$ contributions (97.3 \%) along with Sm 5$d$ (0.46 \%) and B 2$p$ contributions ($\sim$ 2\%) at $E_F$. There is a peak at around -1 eV, which is in good agreement with the $x$-ray photoemission spectra (XPS) \cite{Bucher} and the ARPES spectra \cite{Jiang,Denlinger,Biswas}.

In Fig. \ref{ggaUeq7dos}(b), we show Sm 5$d$, B 2$p$ and Sm 4$f$ PDOS together where Sm 4$f$ contribution is rescaled by 10 times for better comparison. It is clear that Sm 4$f$ - B 2$p$ mixing is stronger in this case compared to the results found in GGA calculations shown in Fig. \ref{ggados}. This is expected as the consideration of electron correlation brings the Sm 4$f$ energies in closer proximity to B 2$p$ energies. While Sm 5$d$ states with $t_{2g}$ symmetry hybridize strongly with the B 2$p$ states, 5$d$ states with $e_g$ symmetry appear to mix with 4$f$ states (see Fig. \ref{ggaUeq7dos}(c)).

\begin{figure}
 \includegraphics[scale=.5]{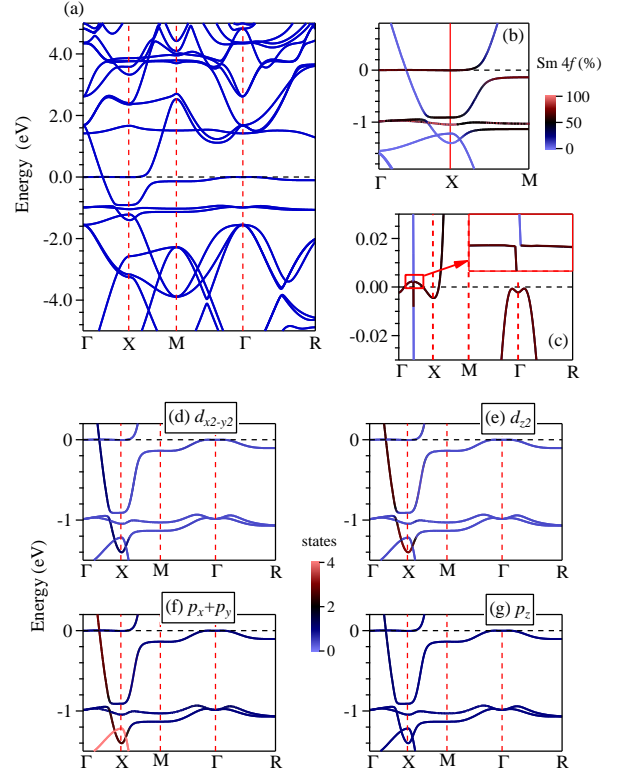}
  \vspace{-1ex}
  \caption{(color online) (a) Band structure calculated using GGA+SOC+$U$ ($U$ = 7 eV, $J$ = 0.83 eV) method. (b) Sm 4$f$ character of the bands near Fermi level are highlighted by the color plots exhibiting signature of band inversion. Weakly dispersive band at -1 eV comes from Sm 4$f$ states. (c) Near Fermi level region in an expanded energy scale to see the band crossing. The color plot exhibit Sm 4$f$ character of the bands crossing the Fermi level. The band characters in the color plot by projecting the eigenstates onto (d) Sm 5$d_{x^2-y^2}$, (e) Sm 5$d_{z^2}$, (f) B (2$p_x$+2$p_y$) and (g) B 2$p_z$ states.}
   \label{ggaUeq7band}
\end{figure}

The electronic band structure using GGA+SOC+$U$ is shown in Fig. \ref{ggaUeq7band}. The flat bands due to Sm 4$f_{5/2}$ states are observed near the Fermi level and near -1 eV energy. A highly dispersive band cross the Fermi level along $\Gamma X$. A gap opens up at the Fermi level along $XM$ direction due to the hybridization of Sm 4$f$ states with the highly dispersive states having dominant B (2$p_x$+2$p_y$) character along with some Sm 5$d_{z^2}$ character as evident from the plots in Figs. \ref{ggaUeq7band}(d) - \ref{ggaUeq7band}(g).
Based on DFT+$U$ calculations using the VASP code, Chang {\it et al.} have shown that the hybridization gap in SmB$_6$ between the localized Sm 4$f$ bands and the conduction bands remains almost unchanged as $U$ is varied up to values as large as 8 eV \cite{Chang}. However, we observe significant influence of $U$ and consideration of exchange correlation potential on the hybridization gap as can be anticipated in such a strongly correlated system. The band inversion like scenario near the Fermi level observed here is consistent with the conclusions from the experimental results.

The highly dispersive band crossing the Fermi level along $\Gamma X$ line possesses dominant B 2$p_x$+2$p_y$ character and hybridizes with the Sm 4$f$ band. The data in Fig. \ref{ggaUeq7band}(c) exhibit an electron pocket primarily formed by the Sm 4$f$ states around $X$ point. A hole like bubble is observed along the $\Gamma X$ line. No other band crossing is found in the data.

The flat bands around -1 eV appear due to correlation induced effects among 4$f$ electrons and can be attributed to the signature of the unscreened feature (incoherent feature) in the spectral functions. The band crossing at $X$-point around -1.5 eV energy gets significantly modified due to strong enhancement of Sm 4$f$ - B 2$p$ hybridizations with the increase in $U$. Therefore, the bands shown in Fig. \ref{ggaUeq7band} possess strong mixed character.

\begin{figure}
   \includegraphics[scale=.45]{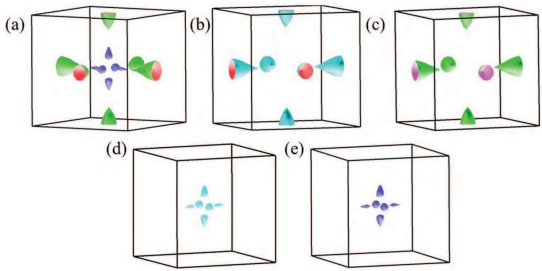}
    \vspace{-64ex}
  \caption{(color online) (a) Fermi surface using GGA+SOC+$U$ method ($U$ = 7 eV, $J$ = 0.83 eV). (b) and (c) Electron pockets around $X$ point. (d) and (e) hole pockets formed around $\Gamma X$ vector. All these Fermi pockets possess dominant 4$f$ character.}
  \label{FSggaUeq7}
\end{figure}

The scenario discussed above in the band dispersion plots are well manifested in the Fermi surface shown in Fig. \ref{FSggaUeq7}. The Fermi surface consists of four distinct Fermi sheets; two of them are due to the electron pockets formed around the $X$ points and two are due to the hole pockets formed on the $X-\Gamma$ line. These results matches well with the high frequency quantum oscillations results \cite{Tan,Hartstein}. While many recent studies suggest that dynamical mean field theory (DMFT) provides a better description of the electronic structure of correlated systems, Kim {\it et al.} \cite{Junwon} have performed the band structure calculations of SmB$_6$ using both DFT and DMFT methods, and concluded that the band structure obtained by using DMFT and DFT near the Fermi level are essentially the same. Here, we show that our calculated results are consistent with the experimental findings, which is remarkable. We believe that B2$p$-Sm4$f$ hybridization plays a key role in the electronic structure in this system, which could be captured well by DFT using suitable choice of interaction parameters. The robustness of the metallic ground states is evident from the fact that all the calculations performed with or without electron-electron Coulomb repulsion, the Fermi surfaces survive. It is to be noted here that although the ground state is metallic, the Fermi level is pinned very close to the band edges with a tiny Fermi energy ($\approx$ 4.5 meV). Thus, a small disorder can localize the electrons easily giving rise to an insulating behavior.

It is to note here that many experiments have shown deviation of the electronic properties of SmB$_6$ from Kondo insulating scenario exhibiting it not to be a true insulator. Wakeham \cite{Wakeham} {\it et al.} have argued that for an insulator without any impurity states or magnetic excitations, the only contribution to the specific heat will be from phonons and so there will not be any linear temperature dependence to the specific heat. But the low temperature contribution to the specific heat is linear in temperature for SmB$_6$ \cite{Phelan,Gabani}. Through measurement of the specific heat in the single crystals and a ground powder of SmB$_6$, they showed that the residual linear term in the low temperature specific heat of SmB$_6$ is predominantly a bulk property and does not originate from the conductive surface states \cite{Wakeham}. In another experiment, Laurita {\it et al.} have used time domain terahertz spectroscopy to investigate the low energy optical conductivity within the hybridization gap of single crystals of SmB$_6$ prepared from both optical floating zone and aluminium flux methods. They have also claimed that this material exhibits significant three dimensional bulk conduction band originating within the Kondo gap \cite{Laurita}. All these experimental results further support bulk metallicity in SmB$_6$ as found in our study. Moreover, the finding of the Fermi surface in quantum oscillation measurements by Tan {\it et al.} \cite{Tan} and the ARPES results by Frantzeskakis {\it et al.} \cite{Fran} are consistent with our results providing confidence to the conclusions of our study. While submitting our paper, we notice a publication \cite{HKPal}, which proposed a model requiring small electron doping and strong particle-hole asymmetry to explain the low temperature anomalies in SmB$_6$. Our results show the presence of such states and particle-hole asymmetry in the electronic structure of pristine SmB$_6$.

\section{Conclusions}

In Summary, we have performed detailed electronic structure calculations of SmB$_6$ using different exchange potentials, spin-orbit coupling and electron correlation strength. We find that Sm 4$f$ multiplets calculated using GGA+SOC+$U$ scheme better describes the experimental ARPES data than the results obtained using GGA+SOC, mBJ+SOC and mBJ+SOC+$U$ methods. The electron correlation strength and exchange interaction strength are found to be 7 eV and 0.83 eV, respectively. The ground state of this material is found to be metallic in all the calculations performed. There are small electron pockets around $X$-point and hole pockets around $\Gamma X$ line having dominant 4$f$ character. The size of the hole pockets is very small indicating proximity to Lifshitz type transition, which might be important for the emergent exoticity in this material. In addition, we discover that the B 2$p$ contribution at $E_F$ is about 4.5 times of the Sm 5$d$ contributions. This trend has been observed in all the calculations using GGA, GGA+SOC, mBJ+SOC and mBJ+SOC+$U$. This suggests strong Sm 4$f$-B 2$p$ hybridization in the electronic structure that is presumably responsible for the anomalous transport properties in this material. All these results provide following important conclusions; (i) consideration of both, electron correlation and spin-orbit coupling is important to derive the electronic properties of this system, (ii) the exchange correlation potential treated within GGA method works better to capture the electronic structure of this rare-earth based strongly correlated material, and (iii) the ground state of this material is metallic with small hole and electron pockets that might be responsible for the exotic electronic properties of this material.


\begin{thebibliography}{99}
\expandafter\ifx\csname url\endcsname\relax
  \def\url#1{\texttt{#1}}\fi
\expandafter\ifx\csname urlprefix\endcsname\relax\def\urlprefix{URL }\fi
\providecommand{\bibinfo}[2]{#2}
\providecommand{\eprint}[2][]{\url{#2}}


\bibitem{Fisk}
G. Aeppli and Z. Fisk, Comments. Condens. Matter Phys. {\bf 16}, 155 (1992);
Y. S. Lee, S. J. Moon, S. C. Riggs, M. C. Shapiro, I. R. Fisher, B. W. Fulfer, J. Y. Chan, A. F. Kemper, and D. N. Basov, Phys. Rev. B {\bf 87}, 195143 (2013).

\bibitem{CeRhCoSi3}
S. Patil, K. K. Iyer, K. Maiti, and E. V. Sampathkumaran, Phys. Rev. B {\bf 77}, 094443 (2008); S. Patil, V. R. R. Medicherla, R. S. Singh, S. K. Pandey, E. V. Sampathkumaran, and K. Maiti, Phys. Rev. B {\bf 82}, 104428 (2010).

\bibitem{Rise}
P. Riseborough, Heavy fermion semiconductors. Adv. Phys. {\bf 49}, 257 (2000).

\bibitem{Cole}
P. Coleman, \textit{Handbook of Magnetism and Advanced Magnetic Materials} (Wiley, New York, 2007), Vol. 1, pp. 95–148.

\bibitem{Anto}
V. N. Antonov, L. V. Bekenov, and A. N. Yaresko, Adv. Condens. Matter Phys. {\bf 2011}, 1 (2011).

\bibitem{Anderson}
P. W. Anderson, Phys. Rev. Lett. {\bf 104}, 176403 (2010).

\bibitem{Menth}
A. Menth, E. Buehler, and T. H. Geballe, Phys. Rev. Lett. {\bf 22}, 295 (1969).

\bibitem{Coleman}
M. Dzero, K. Sun, V. Galitski, and P. Coleman, Phys. Rev. Lett. {\bf 104}, 106408 (2010).

\bibitem{Galitski}
M. Dzero, K. Sun, P. Coleman, and V. Galitski, Phys. Rev. B {\bf 85}, 045130 (2012).

\bibitem{Wolgast}
S. Wolgast, \c{C}. Kurdak, K. Sun, J.W. Allen, D. J. Kim, and Z. Fisk, Phys. Rev. B {\bf 88}, 180405(R) (2013).

\bibitem{Kim}
D. J. Kim, J. Xia, and Z. Fisk, Nat. Mater. {\bf 13}, 466 (2014).

\bibitem{DJKim}
D. J. Kim, S. Thomas, T. Grant, J. Botimer, Z. Fisk, and J. Xia, Sci. Rep. {\bf 3}, 3150 (2013).

\bibitem{Neupane}
M. Neupane, N. Alidoust, S. Y. Xu, T. Kondo, Y. Ishida, D. J. Kim, C. Liu, I. Belopolski, Y. J. Jo, T. R. Chang, H. T. Jeng, T. Durakiewicz, L. Balicas, H. Lin, A. Bansil, S. Shin, Z. Fisk, and M. Z. Hasan, Nat. Commun. {\bf 4}, 2991 (2013).

\bibitem{Jiang}
J. Jiang, S. Li, T. Zhang, Z. Sun, F. Chen, Z. R. Ye, M. Xu, Q. Q. Ge, S. Y. Tan, X. H. Niu, M. Xia, B. P. Xie, Y. F. Li, X. H. Chen, H. H. Wen, and D. L. Feng, Nat. Commun. {\bf 4}, 3010 (2013).

\bibitem{NXu}
N. Xu, X. Shi, P. K. Biswas, C. E. Matt, R. S. Dhaka, Y. Huang, N. C. Plumb, M. Radovic, J. H. Dil, E. Pomjakushina, A. Amato, Z. Salman, D. M. Paul, J. Mesot, H. Ding, and M. Shi, Phys. Rev. B {\bf 88}, 121102(R) (2013).

\bibitem{Fran}
E. Frantzeskakis, N. de Jong, B. Zwartsenberg, Y. K. Huang, Y. Pan, X. Zhang, J. X. Zhang, F. X. Zhang, L. H. Bao, O. Tegus, A. Varykhalov, A. de Visser, and M. S. Golden, Phys. Rev. X {\bf 3}, 041024 (2013); S. V. Ramankutty, N. de Jong, Y. K. Huang, B. Zwartsenberg, F. Massee, T. V. Bay, M. S. Golden, and E. Frantzeskakis, J. Elect. Spect. Relat. Phenom. {\bf 208}, 43 (2016).

\bibitem{sparpes}
N. Xu, P. K. Biswas, J. H. Dil, R. S. Dhaka, G. Landolt, S. Muff, C. E. Matt, X. Shi, N. C. Plumb, M. Radovic, E. Pomjakushina, K. Conder, A. Amato, S. V. Borisenko, R. Yu, H. M. Weng, Z. Fang, X. Dai, J. Mesot, H. Ding, and M. Shi, Nat. Commun. {\bf 5}, 4566 (2014).

\bibitem{GLi}
G. Li, Z. Xiang, F. Yu, T. Asaba, B. Lawson, P. Cai, C. Tinsman, A. Berkley, S. Wolgast,
Y. S. Eo, D. J. Kim, C. Kurdak, J. W. Allen, K. Sun, X. H. Chen, Y. Y. Wang, Z. Fisk, and
L. Li, Science {\bf 346}, 1208 (2014).

\bibitem{Radar}
P. Hlawenka, K. Siemensmeyer, E. Weschke, A. Varykhalov, J. Sánchez-Barriga, N. Shitsevalova, A. Dukhnenko, V. Filipov, S. Gabáni, K. Flachbart, O. Rader, and E. D. L. Rienks, Nat. Commun. {\bf 9}, 517 (2018).

\bibitem{Tan}
B. S. Tan, Y. T. Hsu, B. Zeng, M. C. Hatnean, N. Harrison, Z. Zhu, M. Hartstein, M. Kiourlappou, A. Srivastava, M. D. Johannes, T. P. Murphy, J. H. Park, L. Balicas, G. G. Lonzarich, G. Balakrishnan, and S. E. Sebastian, Science {\bf 349}, 287 (2015).

\bibitem{Hartstein}
M. Hartstein, W. H. Toews, Y. T. Hsu, B. Zeng, X. Chen, M. CiomagaHatnean, Q. R. Zhang, S. Nakamura, A. S. Padgett, G. R. Gant, J. Berk, M. K. Kingston, G. H. Zhang,M. K. Chan, S. Yamashita, T. Sakakibara, Y. Takano, J. H. Park, L. Balicas, N. Harrison, N. Shitsevalova, G. Balakrishnan, G. G. Lonzarich, R. W. Hill, M. Sutherland, and S. E. Sebastian, Nature Physics {\bf 14}, 166 (2018).

\bibitem{baskaran}
G. Baskaran, arXiv:1507:03477

\bibitem{senthil}
D. Chowdhury, I. Sodemann, and T. Senthil, Nat. Commun. {\bf 9}, 1766 (2018).

\bibitem{harrison}
N. Harrison, Phys. Rev. Lett. {\bf 121}, 026602 (2018).

\bibitem{Wien}
P. Blaha, K. Schwarz, G. K. H. Madsen, D. Kvasnicka, and J. Luitz, \textit{WIEN2K: An Augmented Plane Wave and Local Orbitals Program for Calculating Crystal Properties} (Vienna University of Technology, Austria, 2001).

\bibitem{Perdew1}
J. P. Perdew, K. Burke, M. Ernzerhof, Phys. Rev. Lett. {\bf 77}, 3865 (1996).

\bibitem{Perdew2}
J. P. Perdew, K. Burke, M. Ernzerhof, Phys. Rev. Lett. {\bf 78}, 1396 (1997).

\bibitem{tran}
F. Tran and P. Blaha, Phys. Rev. Lett. {\bf 102}, 226401 (2009).

\bibitem{singh}
D. J. Singh, Phys. Rev. B {\bf 82}, 205102 (2010).

\bibitem{Marsman}
Y. S. Kim, M. Marsman, G. Kresse, F. Tran, and P. Blaha, Phys. Rev. B {\bf 82}, 205212 (2010).

\bibitem{Zhong}
J. Li, C. He, L. Meng, H. Xiao, C. Tang, X. Wei, J. Kim, N. Kioussis, G. M. Stocks and J. Zhong, Sci. Rep. {\bf 5}, 14115, (2015).

\bibitem{Biswas2}
P. K. Biswas, Z. Salman, T. Neupert, E. Morenzoni, E. Pomjakushina, F. von Rohr, K. Conder, G. Balakrishnan, M. Ciomaga Hatnean, M. R. Lees, D. McK. Paul, A. Schilling, C. Baines, H. Luetkens, R. Khasanov, and A. Amato, Phys. Rev. B {\bf 89}, 161107(R) (2014).

\bibitem{smb6latt}
P. Villars and L. D. Calvert, \textit{Pearson's Handbook of Crystallographic Data for Intermetallic Phases} (ASM International, Materials Park, 1991).

\bibitem{LaB6APL}
V. R. R. Medicherla, S. Patil, R. S. Singh, and K. Maiti, Appl. Phys. Lett. {\bf 90}, 062507 (2007).

\bibitem{CaB6EPL}
K. Maiti, Europhys Lett. {\bf 82}, 67006 (2008). K. Maiti, V. R. R. Medicherla, S. Patil, and R. S. Singh, Phys. Rev. Lett. {\bf 99}, 266401 (2007).

\bibitem{Kang}
C. J. Kang, J. Kim, K. Kim, J. Kang, J. D. Denlinger, and B. I. Min, J. Phys. Soc. Jpn. {\bf 84}, 024722 (2015).

\bibitem{Nickerson}
J. C. Nickerson, R. M. White, K. N. Lee, R. Bachmann, T. H. Geballe, and G.W. Hull Jr., Phys. Rev. B {\bf 3}, 2030 (1971).

\bibitem{Cooley}
J. C. Cooley, M. C. Aronson, A. Lacerda, Z. Fisk, P. C. Canfield, and R. P. Guertin, Phys. Rev. B {\bf 52}, 7322 (1995).

\bibitem{Molnar}
S. von Molnar, T. Theis, A. Benoit, A. Briggs, J. Floquet, J. Ravex, and Z. Fisk, \textit{ Valence Instabilities} (NorthHolland, Amsterdam, 1982), p. 389.

\bibitem{Travaglini}
G. Travaglini and P. Wachter, Phys. Rev. B {\bf 29}, 893 (1984).

\bibitem{Demsar}
J. Demsar, V. K. Thorsmolle, J. L. Sarrao, and A. J. Taylor, Phys. Rev. Lett. {\bf 96}, 037401 (2006).

\bibitem{Dai}
F. Lu, J. Z. Zhao, H. Weng, Z. Fang, and X. Dai, Phys. Rev. Lett. {\bf 110}, 096401 (2013).

\bibitem{allen}
J. W. Allen, L. I. Johansson, I. Lindau, and S. B. Hagstrom, Phys. Rev. B {\bf 21}, 1335 (1980).

\bibitem{schmidt}
J. N. Chazalviel, M. Campagna, G. K. Wertheim, and P. H. Schmidt, Phys. Rev. B {\bf 14}, 4586 (1976).

\bibitem{Deep-band}
D. Biswas and K. Maiti, EPL {\bf 110}, 17001 (2015); D. Biswas, S. Thakur, K. Ali, G. Balakrishnan, and K. Maiti, Scientific Reports {\bf 5}, 10260 (2015).

\bibitem{Junwon}
J. Kim, K. Kim, C. J. Kang, S. Kim, H. C. Choi,1 J. S. Kang, J. D. Denlinger and B. I. Min, Phys. Rev. B {\bf 90}, 075131 (2014).

\bibitem{Bansil}
T. R. Chang, T. Das, P. J. Chen, M. Neupane, S. Y. Xu, M. Z. Hasan, H. Lin, H. T. Jeng, and A. Bansil, Phys. Rev. B {\bf 91}, 155151 (2015).

\bibitem{Fu}
L. Fu, C. L. Kane, and E. J. Mele, Phys. Rev. Lett. {\bf 98}, 106803 (2007).

\bibitem{Kane}
L. Fu and C. L. Kane, Phys. Rev. B {\bf 76}, 045302 (2007).

\bibitem{Denlinger}
J. D. Denlinger, J. W. Allen, J. S. Kang, K. Sun, B. I. Min, D. J. Kim and Z. Fisk, JPS Conf. Proc. {\bf 3}, 017038 (2014).

\bibitem{Biswas}
N. Xu, X. Shi, P. K. Biswas, C. E. Matt, R. S. Dhaka,Y. Huang, N. C. Plumb, M. Radovic, J. H. Dil, E. Pomjakushina, K. Conder, A. Amato, Z. Salman, D. McK. Paul, J. Mesot, H. Ding, and M. Shi, Phys. Rev. B {\bf 88}, 121102(R)(2013).

\bibitem{BR}
A. D. Becke and E. R. Johnson, J. Chem. Phys. {\bf 124}, 221101 (2006).

\bibitem{Armiento}
R. Armiento, S. Kummel, and T. Ko¨rzdo¨rfer, Phys. Rev. B {\bf 77}, 165106 (2008).

\bibitem{hexaborides}
S. Patil, G. Adhikary, G. Balakrishnan and K. Maiti, Appl. Phys. Lett. {\bf 96}, 092106 (2010); {\it ibid.} Solid State Commun. {\bf 151}, 326 (2011); S. Patil, G. Adhikary, G. Balakrishnan, and K. Maiti, J. Phys.: Condens. Matter {\bf 23}, 495601 (2011).

\bibitem{Bucher}
M. Campagna, G. K. Wertheim, and E. Bucher, \textit{Structure and Bonding} (Springer, New York, 1976), Vol. 30, p. 99.

\bibitem{Chang}
T. R. Chang, T. Das, P. J. Chen, M. Neupane, S. Y. Xu, M. Z. Hasan, H. Lin, H. T. Jeng, and A. Bansil, Phys. Rev. B {\bf 91}, 155151 (2015).

\bibitem{Wakeham}
N. Wakeham, P. F. S. Rosa, Y. Q. Wang, M. Kang, Z. Fisk, F. Ronning, and J. D. Thompson, Phys. Rev. B {\bf 94}, 035127 (2016).

\bibitem{Phelan}
W. A. Phelan, S. M. Koohpayeh, P. Cottingham, J. W. Freeland, J. C. Leiner, C. L. Broholm, and T. M. McQueen, Phys. Rev. X {\bf 4}, 031012 (2014).

\bibitem{Gabani}
S. Gabani, K. Flachbart, V. Pavlık, M. Orendac, E. Konovalova, Y. Paderno, and J. Sebek, Czech. J. Phys. {\bf 52}, 279 (2002).

\bibitem{Laurita}
N. J. Laurita, C. M. Morris, S. M. Koohpayeh, P. F. S. Rosa, W. A. Phelan, Z. Fisk,
T. M. McQueen, and N. P. Armitage, Phys. Rev. B {\bf 94}, 165154 (2016).

\bibitem{HKPal}
Hridis K. Pal, Phys. Rev. B {\bf 99}, 045149 (2019).

\end{thebibliography}
\end{document}